\documentclass[sigconf]{acmart}

\usepackage[utf8]{inputenc}
\usepackage[T1]{fontenc} 
\usepackage[utf8]{inputenc} \usepackage[english]{babel} %
\usepackage{microtype}   
\usepackage{url}         

\DeclareUnicodeCharacter{2002}{\enspace}
\setcopyright{none}
\renewcommand\footnotetextcopyrightpermission[1]{}
\usepackage{float} 
\AtBeginDocument{%
\sloppy 
 }

\begin{document}

\title{When Handshakes Tell the Truth: Detecting Web Bad Bots via TLS Fingerprints}

\author{Ghalia Jarad}
\authornote{Both authors contributed equally to this research.}
\email{jarad23@itu.edu.tr}
\orcid{0009-0003-6154-2292}
\affiliation{%
  \institution{Istanbul Technical University}
  \city{Istanbul}
  \country{Turkey}
}

\author{Kemal B\i\c{c}akc\i}
\authornotemark[1]
\email{kemalbicakci@itu.edu.tr}
\orcid{0000-0002-2378-8027}
\affiliation{%
  \institution{Istanbul Technical University}
  \city{Istanbul}
  \country{Turkey}
}

\begin{abstract}
Automated traffic continued to surpass human-generated traffic on the web, and a rising proportion of this automation was explicitly malicious; they increasingly rely on AI so AI augmented evasive bots could pretend to be real users, even solve Captchas and mimic human interaction patterns. 
This work explores a less intrusive, protocol-level method: using TLS fingerprinting with the JA4 technique to tell apart bots from real users. Two gradient-boosted machine learning classifiers (XGBoost and CatBoost) were trained and evaluated on a dataset of real TLS fingerprints (JA4DB) after feature extraction, which derived informative signals from JA4 fingerprints that describe TLS handshake parameters. 
The CatBoost model performed better, achieving an AUC of 0.998 and an F1 score of 0.9734. It was accurate 0.9863 of the time on the test set. The XGBoost model showed almost similar results.
Feature significance analyses identified JA4 components, especially ja4\_b, cipher\_count, and ext\_count, as the most influential on model effectiveness. 
Future research will extend this method to new protocols, such as HTTP/3, and add additional device-fingerprinting features to test how well the system resists advanced bot evasion tactics.
\end{abstract}



\settopmatter{printacmref=false} 
\setcopyright{none}               
\renewcommand\footnotetextcopyrightpermission[1]{} 
\pagestyle{plain}                 

\maketitle

\section{Introduction}
Attackers and fraudsters employ various techniques to circumvent conventional defenses. Methods such as cookie spoofing, fake User-Agent strings, automated bots, scraping tools, and VPNs or proxies are used to obscure identity and generate fraudulent sessions. These tactics undermine the reliability of many higher layer signals for bot detection; on the other hand, protocol-layer TLS fingerprints are difficult to manipulate and provide a privacy-preserving approach to prioritizing suspicious connections for further investigation \cite{FoxIO2025}.
Transport Layer Security (TLS) fingerprinting is a passive network analysis technique used to identify or classify client and server software based on observable characteristics of their TLS handshakes. Rather than decrypting the encrypted session payload, the method turns visible parts of unencrypted handshake metadata, primarily the Client Hello messages, such as cipher lists, extension sets, and their order, ALPN, and version preferences, into compact signatures like JA3 or JA4, which are called a “fingerprint” that represents the implementation and configuration choices of a TLS stack not what it sends.
This work investigates whether the characteristics observed in the TLS ClientHello packet during the handshake can distinguish malicious bots from human traffic. For this purpose, it examines how well machine learning techniques capture these differences and tests the use of JA4 fingerprints for web bot detection on a real dataset, besides a practical discussion of its limits.

\section{Related Work}
Web bot detection seeks to improve accuracy while enhancing user experience, CAPTCHA which stands for Completely Automated Public Turing test to Tell Computers and Humans Apart, remains the most popular choice for bot detection. Although CAPTCHA is simple and effective against basic bots, it can interrupt real users, and recent advancements in AI have enabled bots to solve CAPTCHA with impressive accuracy \cite{Bursztein2011,Sivakorn2016, VonAhn2003,Xu2020}.
Another method of bot detection involves the use of crawling traps, like those by \cite{Chen2020,David2021}, referred to as honeypots. This method is designed to differentiate between human visitors and web bots by creating specific links that only bots would follow, such as hyperlinks that match the background color. These strategies are quite effective, but they can be easily navigated around if the creators of web bots are aware of them (or can detect them by analyzing relevant web pages) \cite{David2021}.

After mouse movements and keystroke dynamics achieved great success in identity authentication, researchers looked into their application for bot detection \cite{Acien2020,Chu2018,Iliou2021,Iliou2021b,MartnezLlamas2025,Niu2021,Niu2024,Sharma2025, Dhakal2018, DeAlcala2022}. Despite the significant benefits of these methods in bot detection, they also have a number of drawbacks and difficulties, particularly as more complex evasion techniques become available \cite{Venugopalan2025, AminAzad2020, Datadome2020}. Advanced bots are now using machine learning to
mimic human behavior, making detection increasingly difficult. These bots produced realistic synthetic behaviors by employing  Generative Adversarial Networks (GANs), statistical techniques, and diffusion models \cite{See2024,See2024Westphalkeystroke}. The generated data can then be used by these bots to more realistically simulate human behavior \cite{Acien2020,Iliou2021b}. 
Consequently, identifying these web bots is still a complex task, even when detection systems know of their strategy \cite{Acien2020,Iliou2021b}. 
Furthermore, there are significant vulnerabilities in the data collection of biometric behaviors due to the extensive reliance on JavaScript to track mouse movements and keystroke dynamics, which raises serious security concerns. If a malicious actor stops the script from running \cite{Iliou2021}, it can limit data collection efforts.  Although most researchers report great laboratory results in behavior based bot detection, to date, there is no explicitly documented evidence in the literature that mouse movement or keystroke-based web bot detection methods have been evaluated on truly in-the-wild bot traffic. The majority of studies investigate detection efficacy using bots that are either tested against bots designed to mimic human like behavior in a controlled setting or constructed from known bot frameworks \cite{Acien2020,Folch2023}.
Recent studies suggest that by combining various signals such as the subtle nuances of mouse movements with detailed web logs,we can greatly decrease false positives and bolster defenses against highly sophisticated bots that cleverly seek to avoid detection \cite{Iliou2021,See2024}.
A bot may be able to avoid detection in one way, but it is much more difficult to appear human-like in several behavioral aspects at once. 

TLS fingerprinting is like listening to the way a client and a server greet each other, providing insight into their initial interaction rather than decoding their subsequent messages, so in their handshake process, both sides make small, repeatable choices, such as selecting cipher lists, ordering extensions, and specifying ALPN (Application-Layer Protocol Negotiation) preferences, that reveal how their TLS stacks are configured without requiring decryption of the traffic \cite{Husk2016}. One of these TLS fingerprinting methods is JA3, which creates a short signature from a ClientHello message that encapsulates the typical greeting of an application, while server-side methods like JA3S and JA4+ do the same for ServerHello responses \cite{FoxIO2024}. 
Crucially, JA4+ strengthens this approach by taking into account that a server’s response depends on the Client Hello it receives: the same server may produce different Server Hellos for different clients, but it will reply consistently to the same client every time, so pairing client and server fingerprints raises confidence when attributing sessions to particular applications or threat families while keeping payloads and session privacy intact \cite{FoxIO2024}.
Recent research has investigated new methods for detecting harmful online activities by analyzing specific connection patterns known as JA3S fingerprints combined with machine learning techniques to spot malicious servers \cite{Theofanous2024}.
One of the benefits of using handshake fingerprints is that they provide a more reliable way to link specific type of fingerprints to particular software or malware \cite{Heino2023}, on the contrary of traditional identifiers, such as IP addresses, which can frequently change or be faked, these fingerprints remain consistent even if the connection is established from different locations or through various IP addresses.

\section{TLS Fingerprinting}
Transport Layer Security (TLS) is the standard protocol that protects most internet traffic today \cite{Google2024}, and its current stable version is TLS 1.3. A TLS connection begins with a cleartext handshake: the client sends a TLS Client Hello that advertises supported protocol versions, cipher suites, and extensions, and the server replies with a TLS Server Hello that selects options from the client’s list. Because those Hello messages are exchanged before encryption begins, the Client Hello reveals repeatable implementation choices tied to the application or underlying libraries, and the Server Hello is shaped by the Client Hello it receives. After the negotiation phase finishes and keys are derived, the connection moves to the encrypted phase in which application data flows using the negotiated algorithms and keys. 
Transport Layer Security (TLS) fingerprinting is a passive network analysis technique used to identify or classify client and server software based on observable characteristics of their TLS handshakes. Rather than decrypting the encrypted session payload, the method analyses unencrypted handshake metadata, primarily the Client Hello and Server Hello messages, to derive a unique “fingerprint” that represents the implementation and configuration of a TLS stack.
\subsection{TLS Fingerprinting Methods}
Transport Layer Security fingerprinting captures repeatable, visible choices made during the TLS handshake to identify client and server software without decrypting traffic. Client-side JA3 creates a compact signature from fields in the Client Hello so it recognizes an application’s consistent “way of saying hello.” Server-side approaches like JA3S and the newer JA4+ record how servers reply in the Server Hello and related messages\cite{JohnAlthouse2023,JohnAlthouse2025}, improving accuracy by matching server behavior to the client that invoked it \cite{FoxIO2025}. These techniques were developed because common identifiers such as IP addresses, domains, or certificates easily change or are spoofed; instead of depending on those shifting values, fingerprinting turns the stable, repeatable handshake choices into short signatures that let analysts spot the same program, library, or malware even when it connects to different sites or uses different certificates.

\subsection{TLS Fingerprinting For Distinguishing Between Humans and Bots}
 Bots often exhibit TLS configuration that deviate from those of legitimate users which makes it easy to distinguish them through analysis of TLS fingerprints, these unusual patterns appear as follows:
 \begin{itemize}
 \item Attacks and fraud use many tricks that break conventional defenses: cookie spoofing and proxy chains hide identity, try to fake User-Agent strings, bots and scraping tools generate fraudulent sessions, and VPNs/proxies mask origin. These behaviors make many higher-layer signals unreliable; by contrast, protocol-layer TLS fingerprints are hard to fake at scale and offer a privacy-preserving way to prioritize suspicious connections for investigation\cite{FoxIO2025}.
\item Automation frameworks and bot tools can show up in TLS handshakes as many lightweight scrapers and agents (Scrapy) use different TLS libraries or configurations and therefore emit distinctive Client Hello patterns—specific cipher lists, extension sets, ordering. Those low level, implementation dependent artifacts are reproducible and can be summarized into compact fingerprints such as JA3 \cite{JohnAlthouse2023}, for example, linked a single JA3 value to many Dyre malware samples, showing this method can group related malware even when IPs, domains, or certificates change \cite{FoxIO2025}.
\item Browsers follow consistent TLS habits—stable cipher preferences, characteristic extension order, and current protocol versions—whereas automated or malicious clients often advertise outdated ciphers, deprecated protocols (such as SSL 3.0), or odd extension orders. Those handshake anomalies, combined with client–server fingerprint pairs, provide a clear, privacy‑preserving method, actionable indicators of non‑human traffic suitable for modern bot-detection pipelines\cite{Husk2016}.
\item Legitimate browsers show internal consistency: the User‑Agent string (application, version, OS, device class) normally matches the TLS Client Hello characteristics observed—extension sets, cipher‑suite order and protocol version. When a client claims to be “Chrome on iOS” but its TLS handshake reflects a different library or cipher ordering, that inconsistency is a strong sign of impersonation or automation \cite{Husk2016}.
Empirical work \cite{Aneja2024} shows that different browsers produce noticeably different TLS message sequences and lengths for the same page, yielding roughly 30–35\% dissimilarity across browsers; they attribute much of this variation to differences in the cipher‑suite list and resulting handshake behavior. In practice, this means a spoofed User‑Agent backed by a mismatched Client Hello is detectable with high confidence by comparing expected TLS patterns for the claimed platform against the observed handshake to flag impersonation, bots, or other spoofing attempts reliably.
\item JA4H\_ab is an application-level fingerprint \cite{FoxIO2025} that ties together TLS characteristics and HTTP request features for a specific HTTP method: it records the TLS client hello profile (cipher suites, extensions, version preferences) together with which HTTP headers are present, their order, and characteristic values when a client issues that particular method, producing a compact signature unique to that application and usage pattern. Because libraries and runtimes (for example Python, Java) tend to generate consistent TLS and header shapes, JA4H\_ab values cluster by implementation, while custom clients (VPNs, game clients, desktop apps) produce distinct JA4H\_ab signatures. A straightforward, practical heuristic for spotting non‑human traffic is the absence of an Accept‑Language header in the HTTP portion of the JA4H\_ab fingerprint: interactive browsers and human-driven clients normally include Accept‑Language to express locale preferences, so its omission is a strong signal that the client was scripted or automated rather than human‑operated.
\end{itemize}
\section{Methodology}
In this study, we are applying practical machine learning methods to see if we can effectively distinguish between harmful automated traffic came from web bots and genuine human traffic using TLS fingerprints in the JA4 format, since manipulating these fingerprints is hard for bots, even if they use various tactics like changing their IP, it will not be effective to evade detection.
The strategy capitalizes on the sophisticated challenges that bots face in their ongoing attempts to deploy evasion techniques such as IP rotation. This complexity positions JA4 fingerprints as a powerful and trusted signal for identifying bot activity. 
\subsection{Threat Model}
This work protects against bots by identifying client software through JA4 TLS fingerprints, and therefore succeeds when adversaries rely on non-browser network stacks or make only apparent evasions; specifically, it reliably detects the following kinds of bots:
\begin{itemize}
    \item \texttt{Automated scripts and Scrapers:} The method effectively detects automated scripts (e.g., Python `requests`, standard `curl`) that do not attempt to mask their TLS handshake. These tools have distinct fingerprints that differ significantly from standard web browsers (Chrome, Firefox, Safari). 

    \item \texttt{Header Spoofing:} It protects against bots that rotate "User-Agent" strings or IP addresses. Since JA4 fingerprints the underlying cryptographic stack (TLS Client Hello), simply changing the User-Agent header or using a proxy does not change the JA4 fingerprint, allowing the defender to identify the bot despite these evasions. 
    \item Malicious bots that use non-standard TLS libraries or configurations which result in unique, known-bad fingerprints (e.g., C2 beacons, specific vulnerability scanners). 
\end{itemize}
The boundary of protection is equally important: the method cannot distinguish a human from an automated actor in following cases: 
\begin{itemize}
    \item \texttt{Full Stack Emulation:}  that uses a real browser engine (Puppeteer/Playwright/Selenium driving an actual Chrome binary). Because the bot is using the real browser's network stack, the JA4 fingerprint is identical to a legitimate user \cite{Selenium2004}. 
    \item \texttt{Advanced TLS Spoofing:}  it is also ineffective against sophisticated adversaries using specialized libraries designed to mimic browser fingerprints. These tools intentionally construct TLS Client Hello packets to bitwise-match legitimate browser fingerprints. 
\end{itemize}
In short, as shown in Table~\ref{tab:ja4-protection}, JA4 is a robust signal for client application identification and opportunistic bot mitigation, but it is not a standalone authentication or attribution mechanism and must be combined with other signals when facing full stack emulation or advanced TLS spoofing.

\begin{table*}[ht]
  \caption{JA4 Protection Capabilities Against Bot and Spoofing Activities}
  \label{tab:ja4-protection}
  \begin{tabular}{p{6cm}p{3cm}p{7cm}}
    \toprule
    \textbf{Feature of Bot} & \textbf{JA4 Protection Capability} & \textbf{Explanation} \\
    \midrule
    Automated Scripts and Web Scrapers & High & Distinct TLS stacks vs browsers. \\
    Header Spoofing (User-Agent/IP Rotation) & High & JA4 uses TLS handshake, not headers. \\
    Full Stack Emulation via Browser Automation & Low / None & Share the same TLS stack as real browser. \\
    Sophisticated TLS Fingerprint Spoofing & Medium & Specialized tools can mimic browser TLS handshakes. \\
    \bottomrule
  \end{tabular}
\end{table*}

\subsection{Dataset Description and Analysis}

The dataset used in this study, JA4DB \cite{FoxIO2024}, is a community-maintained repository of JA4 fingerprints collected through two primary strategies: active submission and passive network monitoring.
1.	Active submission (controlled environment): Researchers and developers produce labeled fingerprints by running JA4+ scanner tools against isolated, well-characterized client instances (for example, specific browser builds, bot scripts, or scraping tool samples) in laboratory conditions, so these controlled captures supply the ground-truth labels required for later real traffic monitoring  and collection.
2.	Passive Network Monitoring (Real-World Data Observations): Previously collected labeled fingerprints used to observer their appearance in live real traffic when clients send handshake packets to communities’ servers.
The JA4DB dataset reflects both the cryptographic characteristics of client implementations and their actual occurrence in real traffic while preserving the precision of lab-derived labels and anchoring them to real world patterns.

After analysis and validation, the dataset had 227,404 records, covering benign human traffic, well-known benign crawlers, and a large set of malicious web bot traffic.
\subsection{Labeling Dataset}
The main data source for experiments was the JA4DB repository \cite{FoxIO2024}. After initial analysis, the dataset included 227,404 records. These records covered benign human traffic, well-known benign crawlers, and a large group of malicious bot traffic.
The dataset was labeled following a set of consistent guidelines. Traffic records with application or user\_agent\_string features that included well-known crawler identifiers such as Googlebot, Bingbot, or LinkedInBot were classified as “good bots” and excluded from the training data. For the rest of the traffic, entries were marked as “bad bots” if the application field included the term “bot”; otherwise, they were categorized as benign. 
In the dataset, after removing the good bots, there were approximately 50,212 records of bad bots, accounting for 22.08\% of the total, alongside 148,610 records identified as benign, which made up 65.35\%, while the remaining 32,007 good bot records were excluded. The data was then divided into training and test sets in an 80/20 ratio, utilizing a fixed random seed to ensure consistency, and then evaluating various models on this naturally imbalanced dataset, which accurately mirrors the uneven distributions commonly found in real-world traffic scenarios.

\subsection{Feature extraction}
\begin{figure*}[h]
  \centering
\includegraphics[width=\textwidth]{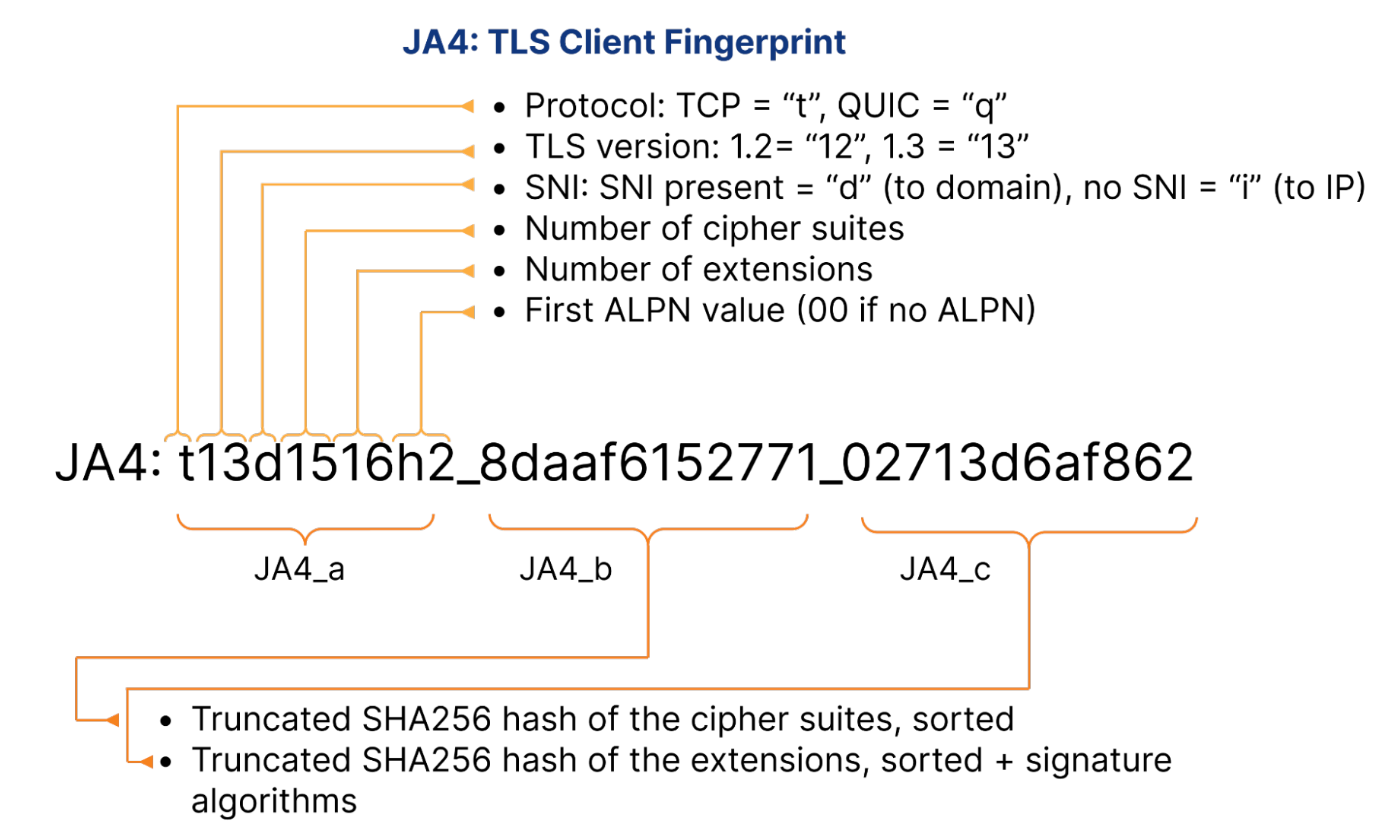}
  \caption{JA4 TLS client fingerprint structure \cite{FoxIO2025}}
  \Description{Parts of JA4 fingerprints and their explanations}
  \label{fig:ja4}
  
\end{figure*}

Raw JA4 strings in the dataset contain useful information that can be parsed and extracted to form structured features for classifier training, as shown in the Figure ~\ref{fig:ja4}. These key features include protocol, TLS version, SNI presence, cipher count, extension count, ALPN code, JA4\_B (cipher suite hash), JA4\_C (extension/signature hash), as well as meta features from the dataset, such as application, OS, device, verified flag, and observation count. By combining these features, the ability to accurately classify regular clients and spot the suspicious bot traffic will be enhanced, by converting the categorical fields within the JA4 string into stable integer codes, the implementation used a specific list of categorical columns, such as protocol, TLS version, ALPN code, ja4\_b hash, ja4\_c hash, application, os, and device. The labelEncoder applies an encoding process to each feature for a more effective and streamlined process.

\section{Models and Experimental Setup}
Two gradient-boosted tree models were trained and compared:
\begin{itemize}
\item {\texttt{XGBoost}}: XGBoost classifier \cite{Chen2016} was trained with 500 boosting trees, a maximum depth of 8, a learning rate of 0.05, a subsample of 0.8, and a colsample of 0.8, with Logloss as the loss function, training used a 80/20 split and a fixed random seed, without synthetic oversampling.
\item{\texttt{CatBoost}}: CatBoost classifier \cite{Prokhorenkova2017} was trained with 500 iterations, a depth of 8, a learning rate of 0.05, Logloss as the loss function, and a fixed random seed, where the categorical features were supplied directly to CatBoost’s Pool data structure so there will be no need for manual encoding and the algorithm could handle them.
\end{itemize}

\section{Findings and Results}
Both models were evaluated on test data using accuracy, precision, recall, F1 score and confusion matrices.
\begin{itemize}
    \item The confusion matrix summarises model predictions by counting true positives, true negatives, false positives, and false negatives; it provides a direct view of how many bad-bot and benign instances were correctly or incorrectly classified and is the foundation for most derived metrics.
    \item Accuracy is the proportion of correctly classified samples (across both classes) and provides a single-number measure of overall performance; while intuitive, it can be misleading on imbalanced datasets because it weights both classes equally, regardless of their importance.
    \item Precision (also called positive predictive value) measures the fraction of predicted bad-bot instances that are actually bad bots; high precision means the model makes few false positive errors.
    \item Recall (also called sensitivity or true positive rate) measures the fraction of actual bad-bot instances that the model correctly identifies; high recall means the model misses few malicious cases.
    \item The F1-score is the harmonic mean of precision and recall and provides a single balanced metric that penalises extreme imbalance between the two; it is especially useful when there is a need to balance avoiding false positives and false negatives.
    \item ROC curve and AUC: The receiver operating characteristic (ROC) curve plots true positive rate versus false positive rate across thresholds, and the area under the ROC curve (AUC) quantifies the model’s ability to rank positive instances higher than negatives; AUC is threshold independent and useful for comparing classifiers, though it can be overly optimistic on highly skewed datasets.
\end{itemize}
\subsection{Results of XGBoost Classifier}
The XGBoost classifier demonstrates strong discriminative performance. Its confusion matrix in Table~\ref{tab:XGBoostConfMatrix} reports 28,699 true negatives and 9,840 true positives, with only 338 false positives and 203 false negatives. The corresponding evaluation metrics in Table~\ref{tab:XGBoostmetrics} are as follows: precision = 0.9668, recall = 0.9798, F1-score = 0.9732, and accuracy = 0.9862. These results indicate that XGBoost reliably classifies benign traffic and maintains a high detection rate for bot instances, with a minimal number of misclassifications.
\begin{table*}
  \caption{Confusion Matrix of XGBoost Classifier}
  \label{tab:XGBoostConfMatrix}
  \begin{tabular}{l r r r}
    \toprule & \textbf{Predicted: Benign} & \textbf{Predicted: Bot} & \textbf{Total} \\ \midrule \textbf{Actual Benign} & True Negative = 28699 & False Positive = 338 & 29037 \\ \textbf{Actual Bot} & False Negative = 203 & True Positive = 9840 & 10043 \\ \midrule \textbf{Total} & 28902 & 10178 & 39080 \\
    \bottomrule
\end{tabular}
\end{table*}

\begin{table}
\centering \caption{Evaluation metrics for XGBoost Classifier} \label{tab:XGBoostmetrics} 
\begin{tabular}{l c c c} 
    \toprule \textbf{Metric} & 
        \textbf{Benign (class 0)} & 
        \textbf{Bot (class 1)} & 
        \textbf{Overall} \\ 
    \midrule Precision & 0.9930 & 0.9668 & - \\ Recall & 0.9884 & 0.9798 & - \\ F1 score & 0.9907 & 0.9732 & - \\ Accuracy & - & - & 0.9862 \\ 
    \bottomrule
\end{tabular} 
\end{table}

\begin{figure}[htbp]
  \centering
\includegraphics[width=\linewidth]{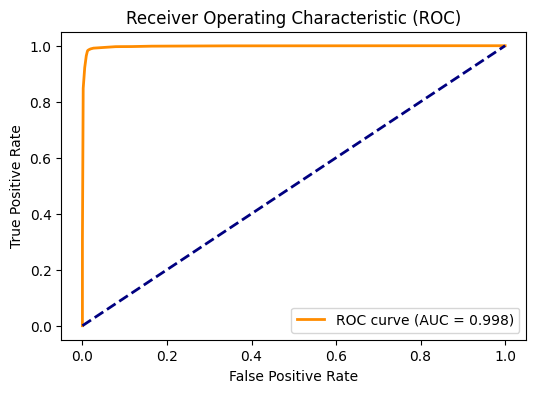}
  \caption{Receiver Operating Characteristic curve for the XGBoost classifier.}
  \Description{Roc Curve of XGBoost Classifier}
  \label{fig:ROCXGBoost}  
\end{figure}

Figure~\ref{fig:ROCXGBoost} presents the Receiver Operating Characteristic (ROC) curve for the XGBoost classifier, where the Area Under the Curve (AUC) is 0.998, which means the model can almost perfectly tell apart bot and benign traffic on the test set. This result shows strong discriminative power for TLS-based fingerprints.
\begin{figure}[htbp]
  \centering
\includegraphics[width=\linewidth]{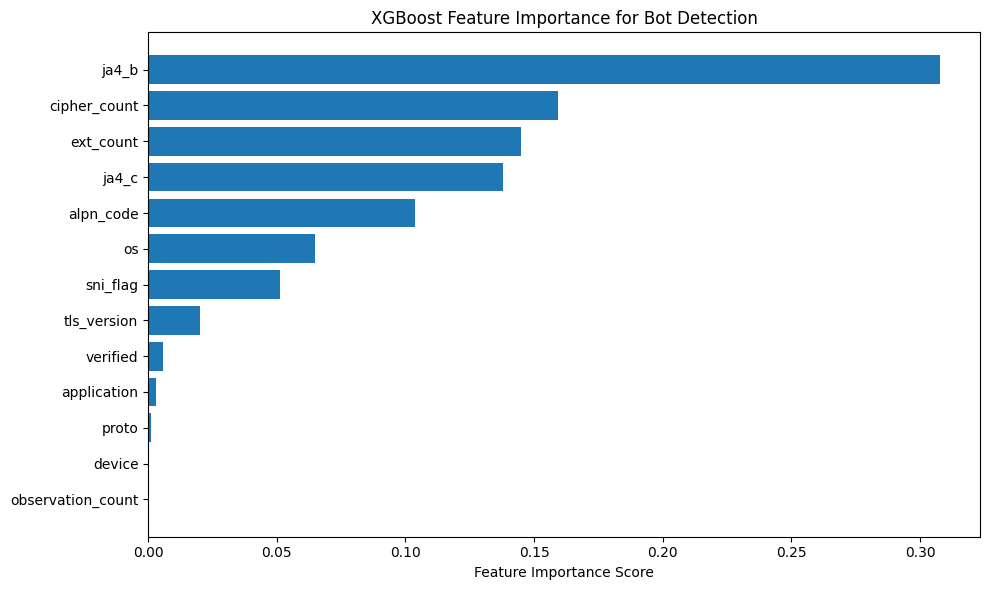}
  \caption{Feature importance for XGBoost Classifier.}
  \Description{Which features have the most effect on classifier results.}
  \label{fig:FeatureImportanceXGBoost}  
\end{figure}
The XGBoost model places the greatest weight on ja4\_b, indicating that this JA4 fingerprint component captures the single most discriminative signal for separating bots from benign clients (See Figure~\ref{fig:FeatureImportanceXGBoost}); immediately following are cipher\_count and ext\_count, which show that the structure and richness of the TLS handshake (number of ciphers and extensions) are strong protocol level indicators of automated versus human traffic.
Medium ranked features such as ja4\_c, alpn\_code, os, sni\_flag, and tls\_version further confirm that subtle variations in TLS composition and client implementation details meaningfully contribute to detection. For example, ALPN\_code (Application-Layer Protocol Negotiation) is useful because bots may use outdated or rare protocols.

\subsection{Results of CatBoost Classifier}
\begin{figure}[htbp]
  \centering
\includegraphics[width=\linewidth]{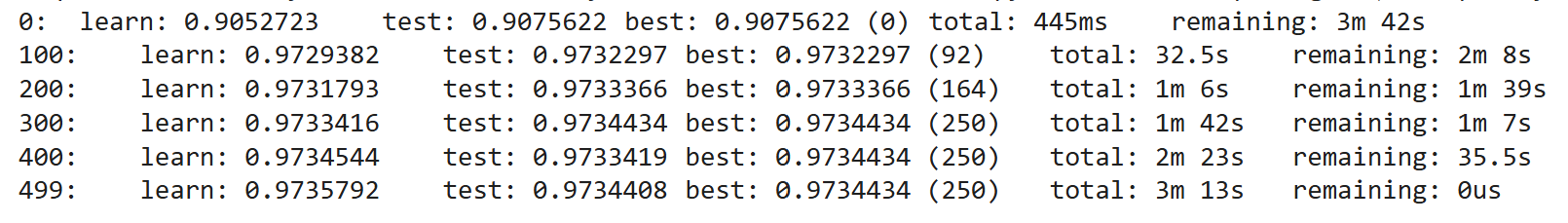}
  \caption{Training log of CatBoost Classifier.}
  \Description{Training Log of CatBoost training.}
  \label{fig:TrainingLogCatBoost}  
\end{figure}

The CatBoost classifier gives almost the same results, but with slight improvements as reflected in training log (Figure~\ref{fig:TrainingLogCatBoost}). 

\begin{table*}
\centering
\caption{Confusion Matrix of CatBoost Classifier}
\label{tab:confusion-matrixCatBoost}
\begin{tabular}{lrrr}
\toprule
 & \textbf{Predicted: Benign} & \textbf{Predicted: Bot} & \textbf{Total} \\
\midrule
\textbf{Actual Benign} & True Negative = 28701 & False Positive = 336 & 29037 \\
\textbf{Actual Bot}    & False Negative = 201  & True Positive = 9842  & 10043 \\
\midrule
\textbf{Total}         & 28902                 & 10178                 & 39080 \\
\bottomrule
\end{tabular}
\end{table*}

\begin{table}
\centering 
    \caption{Evaluation metrics for CatBoost Classifier}
    \label{tab:CatBoostmetricstable} \begin{tabular}{lccc} \toprule \textbf{Metric} & \textbf{Benign (class 0)} & \textbf{Bot (class 1)} & \textbf{Overall} \\ \midrule Precision & 0.9930 & 0.9670 & - \\ Recall & 0.9884 & 0.9800 & - \\ F1 score & 0.9907 & 0.9734 & - \\ Accuracy & - & - & 0.9863 \\ 
\bottomrule 
\end{tabular}
\end{table}

Its confusion matrix (Table~\ref{tab:confusion-matrixCatBoost}) shows 28,701 true negatives, 9,842 true positives, 336 false positives, and 201 false negatives. Table~\ref{tab:CatBoostmetricstable} lists precision at 0.9670, recall at 0.9800, F1-score at 0.9734, and accuracy at 0.9863. CatBoost finds a few more true bots while keeping false alarms just as low, which is clear when comparing the confusion matrices.

\section{Discussion}
\begin{itemize}
\item Both models have low numbers of false negatives and false positives compared to the size of the dataset. CatBoost has slightly fewer false negatives (201 compared to 203) and false positives (336 compared to 338).
\item Precision is almost the same for both models, showing they are equally reliable in positive predictions. CatBoost has a slightly higher recall, so it finds more true positives.
\item The F1 scores and overall accuracy are very close for both models, but CatBoost has a small advantage in both.
\item In practice, either model can be used because both perform well. If it is important to catch as many positive cases as possible without losing precision, CatBoost is slightly better.
\end{itemize}

\section{Conclusion \& Future Work }
This research provides both a conceptual and practical foundation for using protocol based fingerprints, especially TLS fingerprints from the JA4 family, which are practical and reliable for distinguishing malicious bots from real human traffic. By focusing on low-level TLS handshake features that are hard to fake, the study finds that standard gradient boosting classifiers like XGBoost and CatBoost perform well on real-world labeled data. The results support the main idea: JA4-based features reveal patterns that attackers struggle to manipulate or fake, and machine learning models can use these patterns to separate bots from humans with few mistakes.
Also, the dataset and labeling methods rely on the types of bots in the JA4 repository \cite{FoxIO2024}. If attackers adapt their fingerprints or use advanced mimicry, the models may not work as well. So, another important future step could be making adversarial tests against simulated bots that change TLS stacks; to see how hard it is to avoid detection.
The results show that machine learning models using JA4 based features perform well, with high precision and recall, and are resistant to IP address changes by bots. This approach also gives a lasting advantage in the fight against automated, AI-powered bots.

\balance
\bibliographystyle{plainurl}
\bibliography{fullReferences}
\end{document}